# Elastic metamaterial with simultaneously negative Mass Density, Bulk Modulus and Shear Modulus

Ji-En Wu, Xiaoyun Wang, Bing Tang, Zhaojian He[*], Ke Deng[†], and Heping Zhao

[1]Department of Physics, Jishou University, Jishou 416000, Hunan, China

**Abstract:** We present a study of elastic metamaterial that possesses multiple local resonances. We demonstrated that the elastic metamaterial can have simultaneously three negative effective parameters, i.e., negative effective mass, effective bulk modulus and effective shear modulus at a certain frequency range. Through the analysis of the resonant field, it has been elucidated that the three negative parameters are induced by dipolar, monopolar and quadrupolar resonance respectively. The dipolar and monopolar resonances result into the negative band for longitudinal waves, while the dipolar and quadrupolar resonances cause the negative band for transverse waves. The two bands have an overlapping frequency regime. A simultaneously negative refraction for both longitudinal waves and transverse waves has been demonstrated in the system.

Metamaterials (MMs) can be designed to possess many peculiar properties beyond the properties of matter in nature. One famous example is the Electromagnetic (EM) MM with simultaneously negative (effective) permittivity and permeability. Such MMs are often referred as left-handed materials (LHMs), in which EM waves propagate with the wave vectors opposite to the Poynting vectors.[1] LHMs lead to many intriguing phenomena such as the negative refraction[2] and the perfect imaging[3] of EM waves, thus realizing left-handed EM MMs has attracted a great deal of attention in the past two decades and enormous progress has been made along this direction.[4]

Recently, the concept of LHM has been extended to acoustic and elastic medium.[5,6] Extensive efforts have been devoted to realizing acoustic and elastic MMs with negative material parameters.[7-24] Different from EM and acoustic MMs, in an elastic MM, longitudinal and transverse waves usually coexist. Even for the simplest isotropic case, these waves should be described by *three* (rather than *two* in the EM and acoustic cases) independent effective material parameters, i.e., mess density ($\rho_{eff}$), bulk modulus ($\kappa_{eff}$) and shear modulus ($\mu_{eff}$). Actually, the coexisting and coupling of longitudinal and transverse waves in an elastic medium has made the elastodynamics much more complex and richer than acoustics and electromagnetism.[6] From this point of view, a true meaning of elastic LHM requires the realization of triple-negative parameters, which is different from the double-negative requirement for EM and acoustic LHMs.

[*]Corresponding author, e-mail: hezj@jsu.edu.cn
[†]Corresponding author, e-mail: dengke@jsu.edu.cn

Many schemes have been proposed to realize negative parameters in acoustic and elastic MMs. For example, negative mass density can be realized by MMs with building structures of dipolar resonance.[7,8] Negative bulk modulus can be realized by MMs with monopolar-resonance-based building blocks such as the Helmholtz resonators[9] and air bubbles in water.[10] Double negative elastic MMs possesses simultaneously negative bulk modulus and mass density have been achieved by combining two types of structural units with built-in monopolar and dipolar resonances together.[11,12] Moreover, the negative shear modulus has been verified to be related to quadrupolar resonance, and double-negative elastic MMs possesses simultaneously negative shear modulus and negative mass density have also been reported.[13,14] However, to the best of our knowledge, till now few works have been reported on realizing triple-negative elastic MMs, i.e., MMs with simultaneously negative $\rho_{eff}$, $\kappa_{eff}$ and $\mu_{eff}$, and then realizing a negative refraction of longitudinal and transverse waves at the same frequency regime.

In this paper, we designed a two-dimensional elastic MM with multiple resonances. Based on the multiple resonances, we demonstrate that the MM can serve as a negative refraction structure both for longitudinal pressure (P) and transverse (S) waves in a frequency band. The band structure of the system indicated that, at the same frequency regime, for the system there were two hybrid bands with negative dispersion (along ΓX direction) induced by multiple resonances. Further studies elucidate that, one of the hybridized bands possesses simultaneously dipolar and quadrupolar resonances, resulting into the negative $\rho_{eff}$ and $\mu_{eff}$, thus this negative band is only corresponding to S waves. While, another hybrid band possesses simultaneously dipolar and monopolar resonances, inducing the negative $\rho_{eff}$ and $\kappa_{eff}$, hence this negative band is only for P wave. Therefore, simultaneously negative refractions both for P and S waves have been achieved in the overlapping frequency region of these two hybrid bands. All calculations in this paper are fulfilled by the COMSOL Multiphysics.

In order to realize this double negative refraction for P and S waves, we introduce multiple resonances in the two-dimensional structure with square lattice, in which the unit consists of four rod-like scatterers and a cavity among them, as shown in Fig. 1(a). The four rectangular tungsten rods (labeled with blue color) are embedded in a host of foam (dark green). A circular vacuum cavity (white color) is placed in the center. In addition, there are other four rectangular vacuum cavities (white color) next to the four tungsten cylinders, respectively [see Fig. 1(a)]. The lattice constant of square lattice is $a = 6$ mm; the radius of circular vacuum cavity is $r = 0.5$ mm; the rectangular tungsten rods have a size of $L_2 = 1$ mm, $w = 0.6$ mm, and have a distance of $L_1 = 0.75$ mm away from the cavity center. The rectangular vacuum cavities have a size of $L_3 = 0.9$ mm, $w = 0.6$ mm. All of parameters for the materials are shown in the follow [14]: $\lambda = 1.974 \times 10^{11}$ N/m$^2$, $\mu = 1.513 \times 10^{11}$ N/m$^2$, $\rho = 19300$ kg/m$^3$ for tungsten; $\lambda_0 = 6 \times 10^6$ N/m$^2$, $\mu_0 = 3 \times 10^6$ N/m$^2$, $\rho_0 = 115$ kg/m$^3$ for foam, where $\lambda$ and $\mu$ are the Lame constants and $\rho$ represents the mass density.

Band structure of the elastic MM along the $\Gamma X$ direction is shown in Fig. 1(b). We can see that there are two frequency areas with negative refraction. For the lower bands with negative fraction, our further studies elucidate that they are for S waves, which are from quadrupolar resonance and rotation effect of the four tungsten rods, and thus they are out of our interesting. Here, we focus on the two higher negative refraction bands marked by red square and blue triangle in Fig. 1(b) respectively. To be more clearly, we give the zoomed version of these two bands in Fig. 1(c). The upper band ranging from 6.846 kHz to 7.167 kHz is labelled as band A, while the lower one ranging from 6.780 kHz to 7.112 kHz is marked as band B. It is noted that, around 7 kHz, the wavelengths for P and S waves in foam matrix are about 46 mm and 23 mm respectively, which are both much larger than the lattice constant (6mm). Therefore, it can be easily concluded that the two negative bands are not induced by normal Bragg scattering, but they may be originated from the multiple local resonances of the system. In the following, we will focus on the physical mechanism on the two bands.

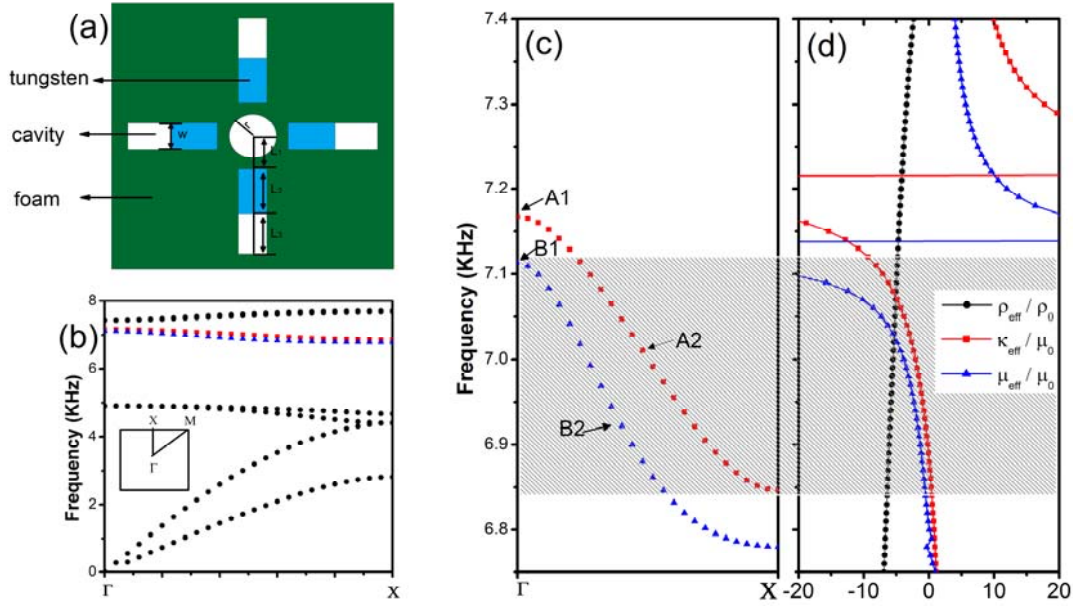

FIG. 1. (a) Unit cell of elastic MM, which is composed of tungsten (blue) rods, foam (dark green) and cavities (white). (b) Band structure of the elastic MM along $\Gamma X$ direction. The inset shows Brillouin zone. (c) Zoomed band structure (6.7 kHz-7.2 kHz) for two colored bands in (b). It can be clearly observed that there are two negative bands labeled as A and B, respectively. (d) Normalized effective bulk modulus ($\kappa_{eff}/\mu_0$), shear modulus ($\mu_{eff}/\mu_0$) and mass density ($\rho_{eff}/\rho_0$) in the concerned frequency. Here, $\rho_0$ and $\mu_0$ are effective mass density and shear modulus of foam matrix, respectively.

Since wavelength is much larger than lattice constant in the concerned frequency band, the system can be described by effective parameters. By employing the method analyzed in Ref. 14 and 15, we calculated effective parameters of the elastic MM as exhibited in Fig. 1(d). We can see that, for the frequency range of band A, the effective mass density and bulk modulus are simultaneously negative. For the frequency range of band B, the effective mass density and shear modulus are simultaneously negative. Hence, it is clearly elucidated that there is an overlapping frequency regime (6.846 kHz to 7.112 kHz) in which the system has three simultaneously negative effective parameters.

To clarify the physical mechanism of these negative parameters, we investigated eigenstates of the two negative bands. We selected the two eigenstates at $\Gamma$ point in band A and band B, which are marked as A1 and B1 respectively in Fig. 1(c). The eigenfields for A1 and B1 are exhibited in Fig. 2(a) and (b) respectively. It can be observed that the A1 eigenstate behaves as a monopolar resonance due to the relative motions of tungsten rods [see Fig. 2(a)], leading to the negative bulk modulus. Whereas, the B1 eigenstate acts as a quadrupolar resonance [see Fig. 2(b)], which is induced by the rotation of tungsten rods around the center cavity, causing the negative shear modulus. By now we have elucidated the origin of negative modulus by the eigenstates at $\Gamma$ point in the band structure, while the mass resonance is almost unable to be observed at this point. In the following, we investigate two eigenstates in the middle of the bands, which are labeled as A2 and B2 in Fig. 1(c) respectively. The corresponding eigenfields are exhibited in Fig. 2(c) and (d). It can be seen that away from the $\Gamma$ point, the pure quadrupolar or monopolar states turn into hybrid states that can be regarded as combinations of a monopolar/quadrupolar state and a dipolar state.[14] For the A2 mode, the hybrid monopolar and dipolar resonances induce the negative band for P waves [as schematically shown in Fig. 2(e)]. While for the B2 mode, the hybrid quadrupolar and dipolar resonances result into the negative band for S waves [as schematically shown in Fig. 2(f)]. Among the two negative bands, there is an overlapping frequency regime (from 6.846 kHz to 7.112 kHz). It indicates that in the overlapping frequency regime the system can act as negative refraction materials both for P waves and S waves.

To further verify the analysis above, we calculate the transmission of the elastic MM in the concerned frequency regime along $\Gamma X$ direction. In the calculation, a MM sample with thickness of nine periods is placed in the foam matrix, as shown in Fig. 3(a). An external normal/tangential harmonic force was exerted on the matrix nearby the left side of MM to act as an input source for P/S waves. The left and right boundaries were terminated with a perfect matched layer (PML) to absorb the outgoing wave. Periodic boundary conditions were imposed on both upper and down sides of the system. The calculations for P and S-wave incidences are shown in Fig. 3(b) and (c) respectively. We can see that the P-wave transmission is high in the frequency regime of band A for P-wave incidence [Fig. 3(b)], while the S-wave transmission is high within the band B for S-wave incidence [Fig. 3(c)]. The results indicate that the two bands are pure and are not hybrid with P waves and S waves, as discussed above. The transmissions are in good agreement with band structures.

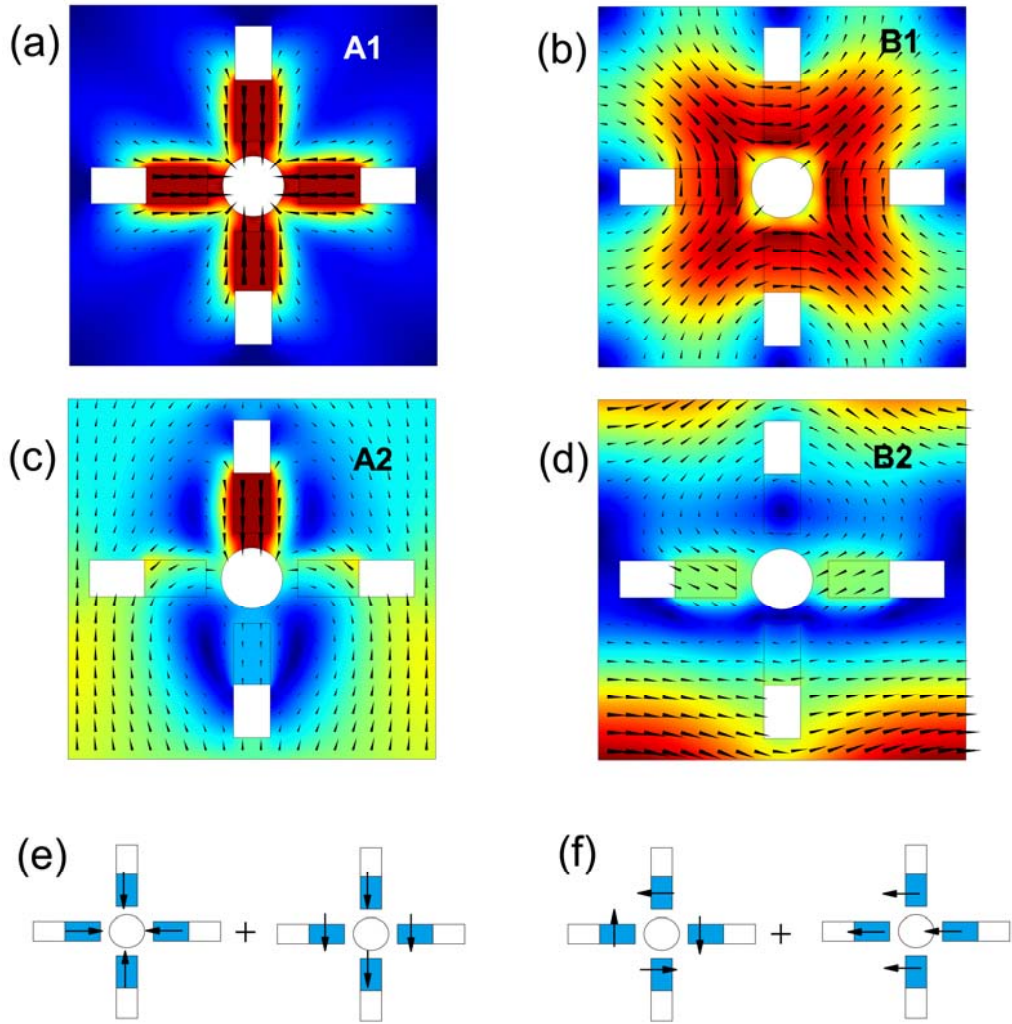

FIG. 2. (a)-(d) Eigenfields for the four eigenstates marked as A1, B1, A2 and B2 in Fig. 1(c), respectively. Here, the triangular arrowheads indicate the field direction, the color represents amplitude (blue for small and red for large). (e) The mode in (c) is schematically shown to arise as a hybridization of a monopolar and a dipolar mode. (f) The mode in (d) is schematically shown to arise as a hybridization of a quadrupolar and a dipolar mode.

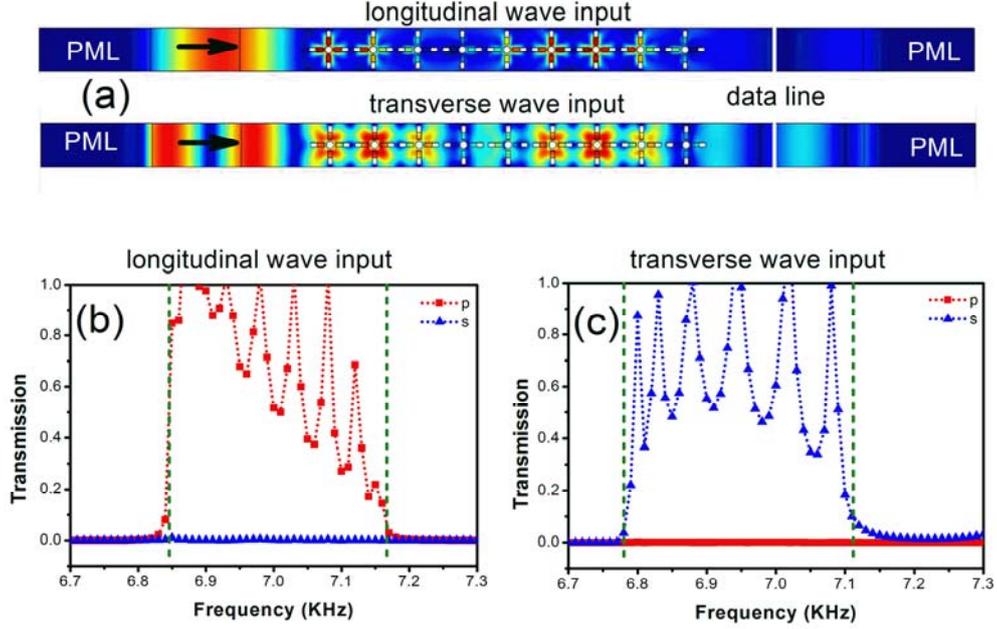

FIG. 3. (a) Model for transmission computation along ΓX direction. (b) The transmission of P and S waves for longitudinal wave input. (c) The transmission of P and S waves for transverse wave input. The two green dotted lines in (b) and (c) indicate the frequency range of band A and B, respectively.

Now we design a $90^0$ wedge sample of the elastic MM to demonstrate the negative refraction effects, as exhibited in Fig. 4(a). The size of the sample is $26a \times 50a$, of which the angle for smaller corner is about $26.6^0$. The two perpendicular surfaces in the wedge are along ΓX direction. The plane source for longitudinal/transverse waves is generated through the normal/tangential harmonic vibration with width 18$a$. The source is normally incident into the wedge along ΓX direction, as schematically shown in Fig. 4(a). Without loss of generality, we choose the frequency as 6.92 kHz. The calculating results for P-waves and S-waves incidence are exhibited in Fig. 4(b) and (c) respectively. In the fields, the color from blue to red indicates the amplitude of the field varying from low to high. It is clearly observed that the incident waves pass through the wedge and are negatively refracted at the interface for both the longitudinal waves and transverse waves input, demonstrating that the negative refraction simultaneously occurs both for P waves and S waves.

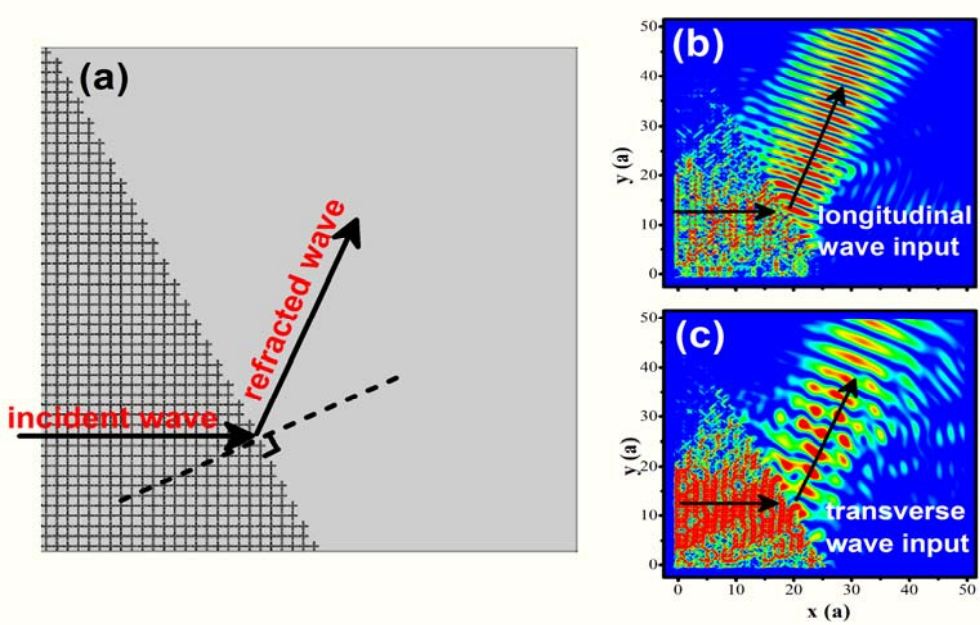

FIG. 4. (a) Schematic illustration for the negative refraction sample. (b, c) Near-field distributions of the sample for longitudinal and transverse waves incidence at frequency 6.92 kHz, respectively.

In order to further study the refraction behavior, we calculate the far field distributions of outgoing waves through the wedge for P-waves and S-waves incidence, which are shown in Fig. 5(a) and (b) respectively. For the case of longitudinal wave incidence, we can see that there is only one refractive beam emitting through the wedge [Fig. 5(a)]. Further analysis indicates that this outgoing beam is S-waves interestingly. For the case of transverse waves incidence, there are two refractive beams emitting through the wedge [Fig. 5(b)]. Through analyzing the refracting angle, the upper beam is P-waves, while the lower beam is S-waves. Moreover, the P-waves refraction is dominant in these two outgoing beams, which is consistent with the results shown in Ref. 13. Here, the total flux ratio of the refracted P waves and S waves at this frequency is approximately *17*. Such unique refraction phenomenon in our system that can convert the wave vibration type of incident waves will have potential application on designing the devices of wave conversion [13].

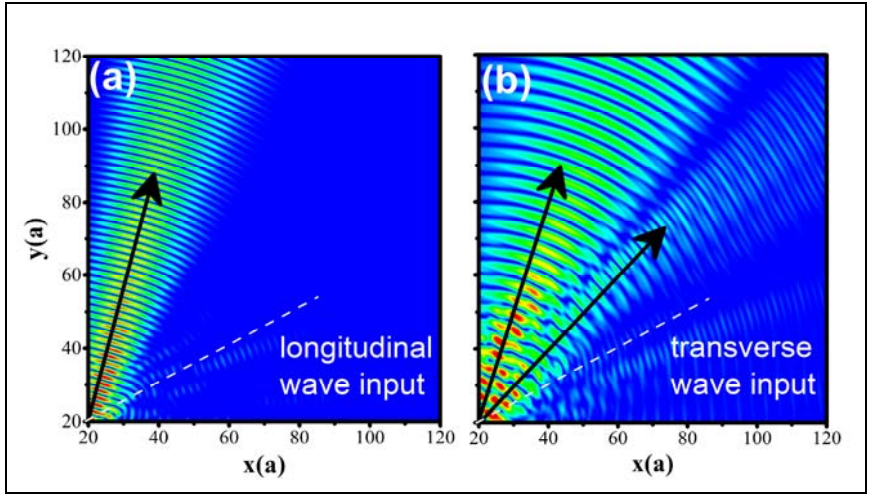

Fig. 5 Far-field distributions of the sample for longitudinal waves incidence (a) and transverse waves incidence (b) at frequency 6.92 kHz, respectively.

In summary, we have designed an elastic metamaterial and demonstrated that three negative effective parameters ($\rho_{eff}$, $\kappa_{eff}$, $\mu_{eff}$) are simultaneously achieved for the designed MM at a certain frequency band. Further analysis elucidated that the negative parameters for $\rho_{eff}$, $\kappa_{eff}$, $\mu_{eff}$ are originated from the dipolar, monopolar and quadrupolar resonances respectively. Two negative bands, one from dipolar and monopolar resonances and the other from dipolar and quadrupolar resonance, are observed in the band structure at the overlapping frequency regime. The negative refraction for both longitudinal and transverse waves was demonstrated by a wedge sample in the overlapping frequency region. Our research for multiple resonant metamaterials could provide a reference for creating and making use of the hybrid resonant modes, and such kind of negatively refracted effects could be utilized in many applications for control elastic wave.

This work is supported by the National Natural Science Foundation of China (Grant Nos. 11464012, 11564012, 11564013, and 11764016), the Natural Science Foundation of Hunan Province, China (Grant No. 2016JJ2100), and the Natural Science Foundation of Education Department of Hunan Province, China (Grant No. 16A170).